\pdfoutput=1 
\documentclass{tufte-handout}
\usepackage{graphicx}
\setkeys{Gin}{width=\linewidth,totalheight=\textheight,keepaspectratio}
\graphicspath{{Figures/}}
\usepackage{booktabs}
\usepackage{units}
\usepackage{fancyvrb}
\fvset{fontsize=\normalsize}
\usepackage{multicol}
\usepackage{lipsum}
\usepackage{natbib}
\def\adx#1:#2\par{\par\halign{\hskip #1##\hfill\cr #2}\par}

\def\rsol{R_\odot}
\def\msol{M_\odot}
\def\lsol{L_\odot}

\def\cp{c_P}
\def\cv{c_V}

\def\nabad{\nabla_{\mathrm{ad}}}

\def\rhoc{\rho_{\mathrm{c}}}
\def\tc{T_{\mathrm{c}}}

\def\diff{{\mathrm d}}

\def\unity{ \hbox{1\kern-.23em l} }
\def\zero{ \hbox{0\kern-.23em |} }
\def\field{ \hbox{I\kern-.23em K} }

\def\braket #1.#2.{\langle #1 \vert #2 \rangle}

\usepackage[fulloldstylenums]{kpfonts} 
\title{Helium Ignition in the Cores of Low-Mass Stars}

\author[]{Alfred Gautschy, CBmA Liestal and ETH-Bibliothek Z\"urich}
\begin{document}

\maketitle

\marginnote[-1.5cm]{This document is best indulged electronically.

                    \bigskip

                    \parindent 0pt
                    CBmA: Center for Basement Asterophysics; an
                    autonomous astronomy research venture of 
                    the same mindset as cbastro.org
                    }

\begin{abstract}
  \noindent In stars with $M_\ast \lesssim 2 \msol$, nuclear burning
  of helium starts under degenerate conditions and, depending on the
  efficiency of neutrino cooling, more or less off-center. The
  behavior of the \emph{centers} of low-mass stars undergoing core
  helium ignition on the $\log\rho - \log T$ plane is not thoroughly
  explained in the textbooks on stellar evolution and the appropriate
  discussions remain scattered throughout the primary research
  literature. Therefore, in the following exposition we collect the
  available knowledge, we make use of computational data obtained with
  the open-source star-modeling package MESA, and we compare
  them with the results in the existing literature. The line of
  presentation follows essentially that of \citet{Thomas1967} who was
  the first who outlined correctly the stellar behavior during the
  off-center helium flashes that lead to central helium burning. The
  exposition does not contain novel research results; it is
  intended to be a pedagogically oriented, edifying compilation of
  pertinent physical aspects which help to \emph{understand} the
  nature of the stars.
\end{abstract}

\newthought{Shortly after Henyey's method came into use} to solve
numerically the equations of stellar structure and evolution, model
stars were followed up the first giant branch and into core helium
burning \citep{Haermschild1964}. In these early computations, helium
ignited in the stars' very centers because neutrino cooling was not
yet accounted for. In his PhD thesis, \citet{Thomas1967}
\sidenote{The ZfA paper is essentially a verbatim reproduction of the thesis.}
computed the evolution of a $1.3 \msol$ star from the ZAMS to the
beginning of core helium burning and, in contrast to the earlier
Schwarzschild \& H\"arm calculations, added neutrino energy
losses. The energy leak in the stellar core induced by the weakly
interacting neutrinos leads to an important, qualitative change in
central stellar structure: Instead of a monotonous temperature drop
from the center, a temperature inversion develops in the deep
interior.  Because of the strong temperature dependence of the nuclear
reaction rates, helium burning starts off-center with the
nuclear-active shell eating slowly its way to the star's center.

The paper of \citet{Demarque1971} contains results from
stellar-evolution computations of low-mass population~II stars ranging
in mass from $0.5$ to $0.85 \msol$. The authors aimed at investigating
the critical core mass at which helium ignites and the subsequent
evolutionary phase with its core cooling. Accordingly, the paper
contains a figure showing, on the $\log\rho - \log T$ plane, the loci
of stellar centers under different physical conditions; in particular,
one plot shows the central cooling and expansion episode during the
initial helium flash. Not the complete evolution from the top of the
giant branch to the arrival of the star model on the horizontal branch
was computed so that the loci on the $\log\rho - \log T$ plane
remained incomplete.
 
At around the same time, \citet{Mengel1981} and \citet{Despain1981}
published papers wherein they painstakingly computed the evolution of
a low-mass Population~II star from the top of the giant branch,
through the helium flashes, onto the horizontal branch (an $0.7 \msol$
star with 8 thermal pulses of the He shell in the Mengel~\&~Demarque
paper and an $0.6 \msol$ star with 12 thermal pulses in the paper of
Despain). \citet{Mengel1981} illustrated the evolution during the
thermal-pulse episode with detailed diagrams for luminosity, and
temperature. A time~--~mass diagram (lately referred to as
Kippenhahn diagram) tracing the extent of convection zones and the
locations of nuclear burning regions documented the closing in on the
stellar center of the thermally unstable helium-burning shell on the
time-scale of the order of a million years. 

\citet{Despain1981} emphasized that his stellar evolution computations
of a $0.5 \msol$ Pop II star did not resort to any artificial shifting
of the thin hydrogen-burning shell in the discretized model star
during its evolution along the giant branch. Therefore, he considered
his computations as a benchmark for older ones to scrutinize their
results, which were obtained with then unproven numerical
simplifications. Despain went on to explain the model star's
structural behavior on its way starting from the ZAMS to the onset of
thermal pulses on the lower AGB. In particular, Despain put forth
physical explanations of numerical findings during the off-center
onset of He burning and the inward evolution of the He-burning
shell. Establishing a local stability analysis; he explained the
thermal instability of the helium shell, which underwent 12 thermal
pulses before core helium burning established itself, and he also
presented a $\log\rho - \log T$ diagram showing the locus of the
star's center and of the temperature maximum during the star's
evolution from the top of the giant branch to the horizontal
branch. Most importantly, Despain indicated that the star's center
moves, due to its expansion and cooling, towards the lower left of the
$\log\rho - \log T$ diagram, whereas the location of maximum
temperature heats at essentially constant density until the
temperature maximum enters a sufficiently low-degeneracy region to
also expand and hence decrease its density and also slightly its
temperature. Together with the \citet{Despain1976} paper, Despain
provided at the time the most thorough discussion and physical
explanations of the mechanical and thermal behavior and the stability
properties of cores of low-mass stars during the onset of helium
burning in their cores. Nonetheless, the loci on the $\log\rho - \log
T$ plane remained fragmentary: Only the initial He-flash phase was
covered.

To the best of the author's knowledge, the numerically established
loci of the low-mass stars' centers on the $\log\rho - \log T$ plane
as already published in \citet{Demarque1971} or \citet{Despain1981}
never made it into textbooks on stellar structure and evolution.  The
{\oe}uvre of \citet{kw}, a benchmark for advanced textbooks on stellar
physics, never discusses the locus of the low-mass star's center
during He ignition; the authors referred to the data
of \citet{Thomas1967} but content themselves with showing the behavior
of the He-ignition mass shell on the $\log\rho - \log T$ plane. Later
in the book, when discussing the evolution of the central region
(\S~33.4), Kippenhahn \& Weigert relied on a plot of \citet{Iben1974}
wherein the loci of the centers of stars with degenerate cores are
plotted \emph{incorrectly} for the phase of helium ignition. Even in the most
recent texts on stellar evolution, the situation has not improved; 
e.g. in \citet[][their Fig.~5.12]{SalarisCassisi2005} the locus of the
$1 \msol $ star's center evolution on the $\log\rho - \log T$ plane
continues to be plotted incorrectly.

Until recently, only a few specially tuned stellar evolution codes
could reliably evolve low-mass model stars up the first giant branch
and the ensuing onset of the helium burning either onto the horizontal
branch or into the clump giant region of the HR Diagram. Following the
ever narrowing hydrogen burning shell during the evolution up along
the giant branch either requires a huge number of timesteps to
accurately follow the evolution of the nuclear burning shell. To
circumvent the problem, elaborate transport prescriptions to shift the
H-burning shell artificially were implemented at the time when
computers were much slower than today, this allowed to maintain big
enough timesteps and hence reduced the computational load
considerably. Furthermore, once helium burning ignites, the timesteps
can easily shrink to a fraction of a day; this after starting the
evolution of the model star with temporal step-widths of the order of
several $10^8$~yrs. Hence, the numerics of stellar evolution codes,
which have to cope with this huge temporal resolution range, needs to
have appropriately robust discretization schemes. Frequently,
shortcuts on to the horizontal branch were taken by the modelers; they
evolved low-mass stars up the giant branch, till shortly before the
onset of He burning. Then, by suitably modifying the numerical models,
the stars were forced onto the horizontal branch so that the full
stellar evolution computations could be resumed there. The paper of
\citet{Serenelli2005} discussed and compared various methods which
were applied in the past.

\smallskip

To gather information on mixing in the deep interior of very
metal-poor low- and intermediate-mass stars, \citet{Suda2010} evolved
their models from the main sequence to the thermally-pulsing AGB.
Suda \& Fujimoto showed a $\log\rhoc - \log\tc$ diagram
with \emph{complete} loci as traced out by the centers low-mass star
models; the properties of the loci were, however, only mentioned en
passant in the text without any deeper explanation. 

Already at the end of the very first computation of a core helium flash,
\citet{Haermschild1964} addressed the question if the ignition episode
could be a dynamical one. Subsequently this aspect generated a
considerable body of literature, a body which continues to grow to the
present. The current opinion arrived at from multidimensional CFD
simulations is that the star remains essentially in hydrostatic
equilibrium during the helium ignition process. The complex and
dynamically changing convection zone in the helium-burning layer
gives, however, rise to potential elemental mixing that cannot be
captured with the simplified and canonically applied mixing-length
model of convection in the conventional stellar-evolution
approaches. The paper of \citet{Mocak2011} gives an impression of the
current state of information and can serve as a guide to the
literature on the fluid dynamical aspects of the ignition of core
helium burning.

As a by-product of the exoplanet search of the \emph{Kepler}
spacecraft mission, a huge number of stars in a selected field in the
sky got highest-precision time-series photometry. Among them are also
red giants exhibiting stochastically driven oscillations whose
frequencies let us probe the otherwise inaccessible internal
structure \citep[see e.g.][]{Bedding2011}.  In particular, the
oscillations whose restoring force have a buoyancy contribution grant
access to the regions that are influenced by nuclear burning;
i.e. these modes can probe a red giant's stage of evolution: they
allow to discriminate between stars being still along the first giant
branch, those being just during the onset of core helium burning, and
the stars already being during central helium burning. It appears that
for the first time we see now observational evidence of stars that are
in the process of igniting helium in their cores
\citep{Bildsten2011}.

The fast present-day personal computers and the advent of a new
generation of stellar evolution codes, in particular the MESA
collection \citep{Paxton2011}, proved to be capable to evolve low-mass
stars into helium core burning even without artificially shifting
narrow H-burning shells during the first giant-branch evolution or
skipping the off-center helium flashes. Hence, it is now possible to
study extensively and in detail even with modest computer-hardware
in\-fra\-structure the behavior of stars during the relatively short-lived
initial off-center helium burning in a shell and its subsequent
evolution toward the star's center.

\newthought{This exposition is going to remind the reader of old
  knowledge}~--~the first of it put forth almost half a century ago.
No new science is going to be unearthed; old facts from the key-papers
referred to above are collected, newly arranged, and freshly
illustrated so that hopefully, after this new iteration, the correct
loci on the $\log\rho - \log T$ plane of the centers of
helium-igniting low-mass stars will eventually find their way into the
upcoming generations of textbooks on stellar physics.

\section{Numerical modeling}
\label{sect:modeling}
All the data to which the ensuing discussion will refer were obtained
with the versatile MESA code suite, which is comprehensively described
in \citet{Paxton2011}.  The computations were performed using release
version 3251. Since we are only interested in the generic behavior of
the model stars, we kept the prescription of the physical ingredients
as simple as possible. Hence, except if mentioned explicitly otherwise
in the text, we computed the stellar-evolution models without
rotation, magnetic fields, and we neglected mass loss.  Convection was
treated according to Henyey's MLT prescription and we adopted the
Schwarzschild criterion for convective instability; the mixing-length
was set ad hoc to 1.8 pressure scale-heights.

Even though different microphysics and in particular more appropriate
treatment of dynamical convection will change the results
quantitatively, probably mainly with respect to abundance profiles and
the associated consequences, we are positive that the following
discussion caught the generic nature of low-mass stars' evolution
through the onset of core helium burning and~--~viewed on the HR
diagram~--~their relatively fast transition from the top of the giant
branch onto the horizontal branch or into the clump on the giant branch.

\section{Low-mass stellar evolution to the onset of helium burning}
\label{sect:low-mass-evolution}
\begin{marginfigure}
	\includegraphics{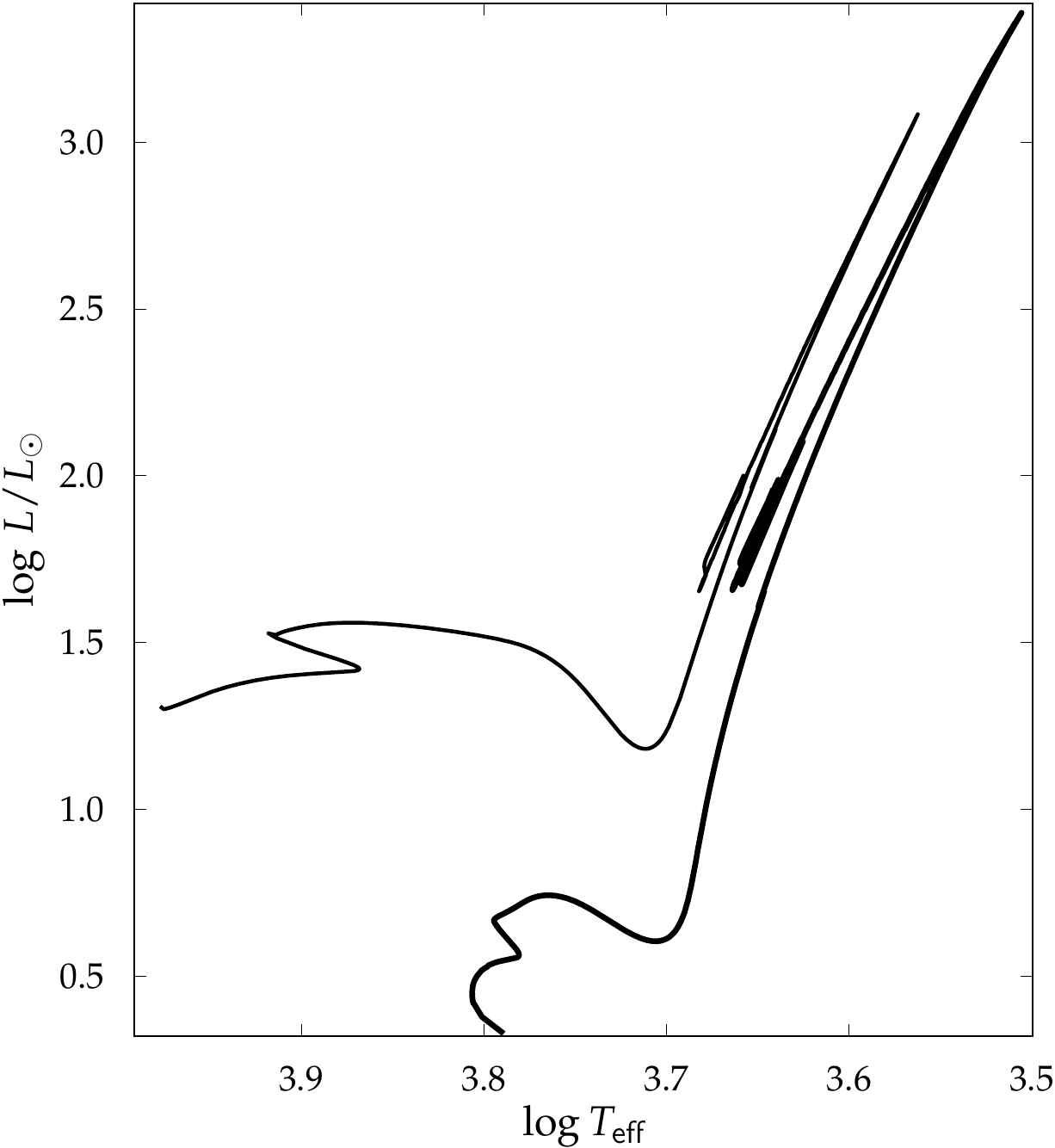} \caption{Conservative evolution
	up to central helium burning, traced out in the HR Diagram of a
	$1.3 \msol$ and a $2.1 \msol$ star, respectively, both
	initially with the abundances $X=0.70,
	Z=0.02$.}  \label{fig:HRD}
\end{marginfigure}

In the following, talk of `low-mass stars' shall refer to stars
whose core helium burning starts under degenerate conditions. Since
neutrino cooling usually causes the maximum of temperature to be
reached off-center, helium burning too starts off-center. Eventually,
the \emph{central} helium-burning stage is reached through a series of
thermal pulses of the inwardly propagating helium-burning shell;
during this process, the centers of low-mass stars trace a generic
path on the density~--~temperature plane, the characteristics of which
lies at the center of attention in this exposition.

Figure~\ref{fig:HRD} shows two representative evolutionary tracks of a
$1.3 \msol$ (heavy line) and a $2.1 \msol$ star (thin line) as
computed with MESA. For the discussion of the off-center onset of
helium burning under degenerate conditions we chose the $1.3 \msol$
case because it was historically the first one for which the
off-center He-flash and the ensuing secondary flashes were presented
\citep{Thomas1967}. The star with $2.1 \msol$ in Fig.~\ref{fig:HRD}
was chosen because its mass just marginally exceeds the critical
stellar mass above which helium burning ignites in the center and
under only weakly degenerate conditions; the difference of the nuclear
evolution is not visible in the track on the HR plane but it is
evident on the $\log\rho_{\mathrm c} - \log T_{\mathrm c}$ plane

The evolution of the model stars, which are discussed in the
following, was computed in quasi-hydrostatic fashion through core
hydrogen burning, and then into central helium burning until central
helium abundance dropped to $Y_{\mathrm c}=1\cdot 10^{-4}$. The
computations started with chemically homogeneous ZAMS models 
assuming $X=0.7, Z=0.02$.

\section{The initial off-center helium flash}
\label{sect:initialflash}
Once a low-mass star reaches the top of the giant branch (e.g. as
shown in Fig.~\ref{fig:HRD}) the maximum temperature is going to
exceed the critical level of about $10^8$~K above which helium starts
to fuse mostly into $^{12}C$ via the $3\alpha$ reaction. The maximum
temperature of the low-mass stars is attained not in the center but at
a mass depth that depends on the total stellar mass and on the
specific microphysics~--~foremost on the efficiency of the neutrino
energy-losses. The particular numbers that will be referred to
throughout this exposition are prone to change depending on the
particular realization of the microphysical processes included in the
evolution computations and on particular choices of the stars. The
qualitative nature of the stellar behavior will, however, remain
unchanged and it is this \emph{qualitative picture} that is the heart
of the present story and that serves us well enough to understand the
stellar behavior.

Table \ref{tab:TP1Selecta} lists a few characteristic stellar
quantities for $1.3 \msol$ star models during the initial helium
flash. The first column lists the model numbers to which the text and
the figures refer to occasionally. The second column contains the age
in years of the model stars counted relative to the maximum of the
initial helium flash, which was encountered at $t_0
=$~4\,522\,060\,464.5~yrs \footnote[][-1.5cm]{The at first sight
  senseless accuracy of such an epoch statement can be justified in
  the context of \emph{relative} timing (i.e. $t^\ast = t-t_0$), which
  is helpful to describe the temporal evolution of the onset of core
  helium burning. For an absolute timing, it is sufficient to remember
  that it takes an $1.3 \msol$ star about $4.5$~Gyrs to the top of the
  first giant branch.}  for the chosen parameters of the evolution
computation. The third column lists the radii of the models, followed
by the total stellar luminosity and the contributions by helium and
hydrogen burning.

\begin{table*}
  \caption[][0.5cm]{Selected global quantities of $1.3 \msol$ models during the 
    initial helium flash
    cycle, which is discussed in detail in the text.}
\label{tab:TP1Selecta}
\centering
\begin{tabular}{c r r r r r}
\hline
\hline
 Model no. & $t^{\ast}/$yrs & $\log R_\ast/\rsol$ & $\log L_\ast/\lsol$ &
$\log L_{\mathrm{He}}/\lsol$ & $\log L_{\mathrm H}/\lsol$ \\
\hline
12\,850 & -5033.59     & 2.204  & 3.38  &  1.41 &  3.38 \\
12\,950 & -0.45        & 2.204  & 3.38  &  6.45 &  3.28 \\
13\,000 & -0.01        & 2.204  & 3.38  &  8.55 &  2.91 \\
13\,025 &  0           & 2.204  & 3.38  &  9.25 &  2.04 \\
13\,050 &  0.01        & 2.204  & 3.38  &  8.85 &  0.88 \\
13\,100 &  0.08        & 2.204  & 3.38  &  7.75 & -0.96 \\
13\,200 &  6.73        & 2.201  & 3.38  &  5.44 & -3.66 \\
13\,500 &  996.71      & 1.823  & 2.84  &  2.96 & -1.33 \\
13\,700 &  9\,542.92   & 1.406  & 2.23  & -0.09 &  0.30 \\
13\,900 &  105\,296.80 & 1.296  & 2.07  & -0.26 & -2.70 \\
13\,984 &  185\,217.60 & 1.304  & 2.07  &  4.23 & -0.30 \\
\hline
\end{tabular}
\end{table*}

\begin{figure}
\includegraphics[width=0.98\textwidth]{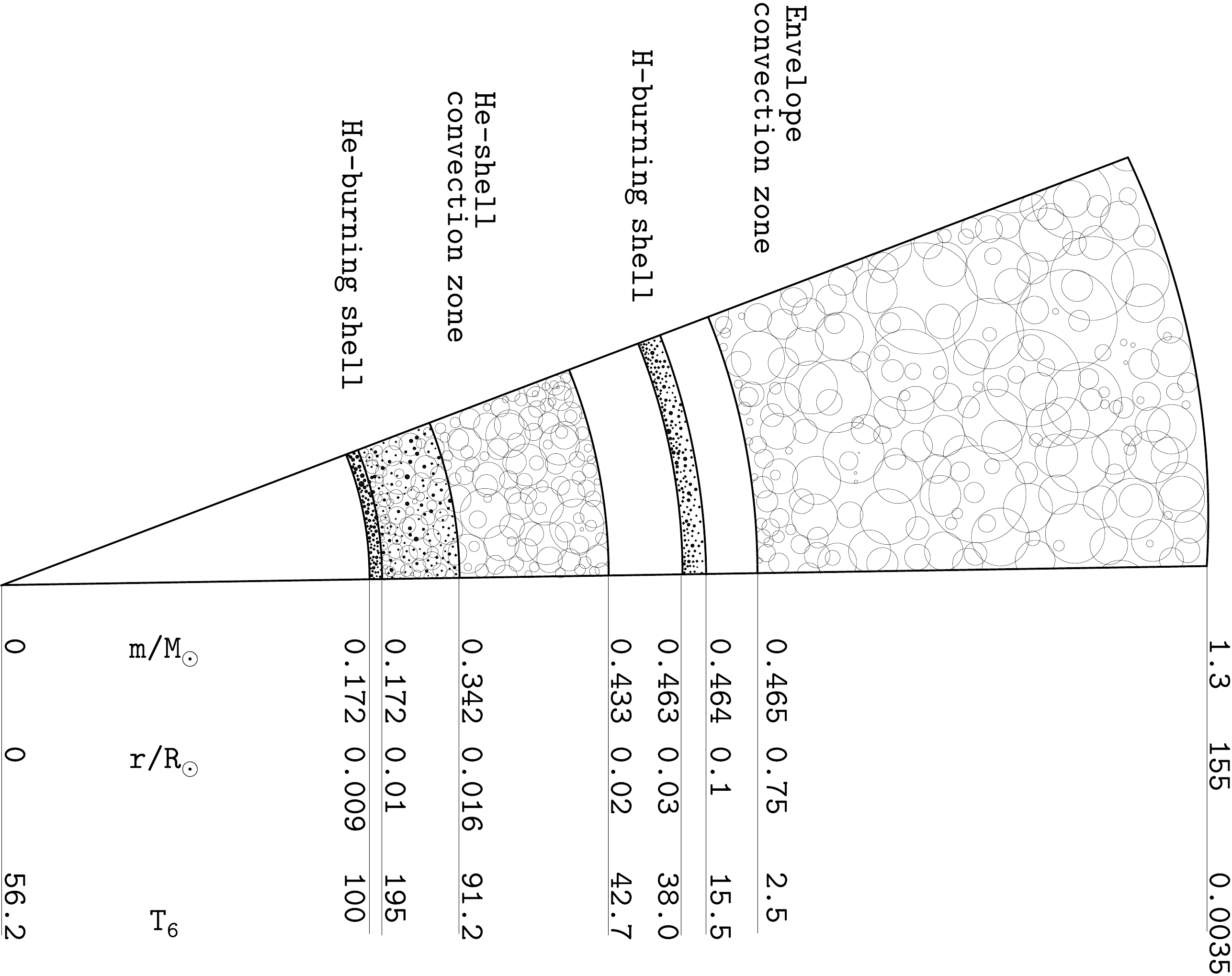}
\caption{Schematic internal structure of a star
model at maximum nuclear energy release during the initial helium
flash.  The scales of the radiative, convective, and nuclear burning
regions are not constant through the sketch. Numerical 
values of selected physical quantities, adopted from model
number 13\,025, are tabulated on the bottom.}
\label{fig:SchematicM013}
\end{figure}

The schematic structure of the stellar model at maximum helium burning
(model no. $13\,025$, at $t^\ast=0$) of the initial helium flash is
sketched and labeled in Fig.~\ref{fig:SchematicM013}. The convective
envelope and the convective shell overlying the helium-burning shell
are hinted at with open circles. Nuclear burning regions, on the other
hand, are hinted at with small black dots. In the graphical
representation, the geometry is not to scale; appropriate physical
quantities at selected locations are listed on the bottom of
the plot, giving the masses, the radii and the corresponding
temperatures, $T_6 \equiv (T/K)/10^6$ at the selected boundaries. A
nuclear burning region was considered as such if the energy generation
rate exceeded $100$~erg/g/s.

\begin{figure}
\includegraphics{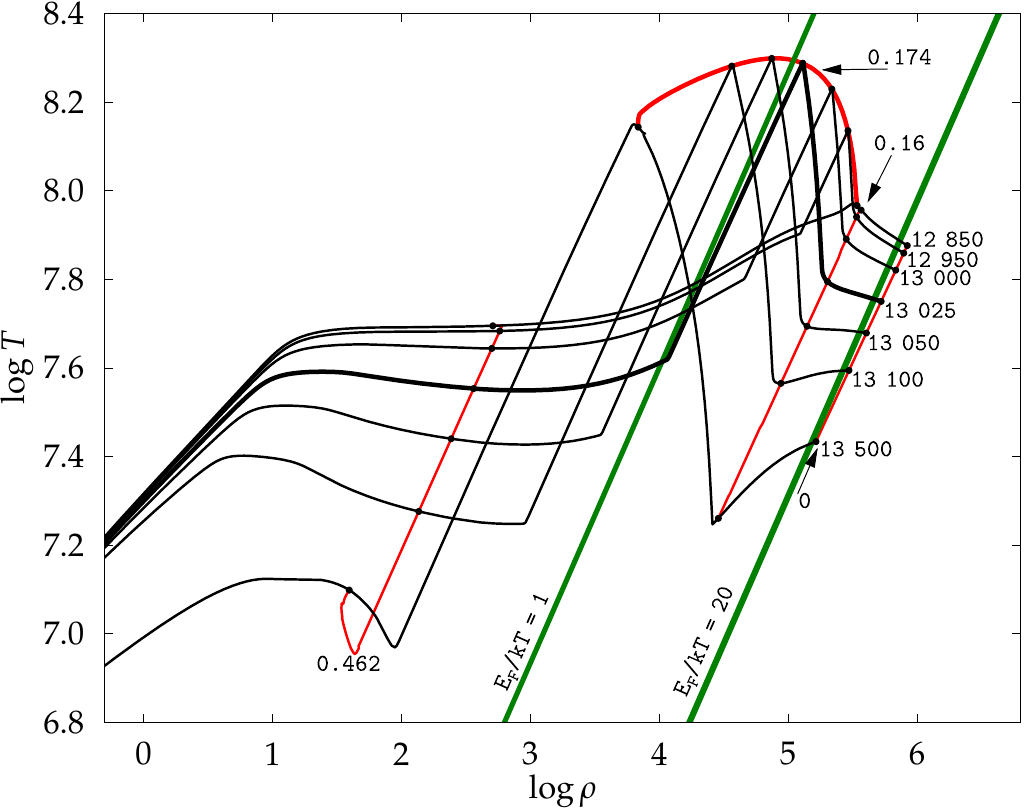}
\caption{Loci on the $\log\rho - \log T$ plane of selected
$1.3 \msol$ models during the early initial He flash (model numbers from
$12\,850$ to $13\,500$). The thicker black line marks the model during
the peak of the initial helium flash.  The red lines trace the
evolution of a few mass shells (their mass depths are measured in
solar-mass units) on the density~--~temperature plane. The heavier
red line~--~at 0.174~--~traces the locus of maximum
$\varepsilon_{\mathrm{nuc}}$ of the helium burning shell during the
early initial flash. 

The green lines with $E_{\mathrm{F}}/kT = 1$ and $20$
illustrate lines of constant degeneracy. Since $E_{\mathrm{F}}\sim
\psi$, with $E_{\mathrm{F}}$ the Fermi energy and $\psi$ the
canonical degeneracy parameter, the low-order approximation of
density in a strongly electron-degenerate environment reads $\rho \sim
\left(\psi k T \right)^{3/2}$ so that $\left.\diff \ln T / \diff \ln
\rho\right\vert_{\psi={\mathrm{const}}} = 2/3$.

N.B. the black lines trace strati\-fications, i.e. the \emph{spatial
  structure} of selected model stars; the red lines delineate
\emph{temporal}, lagrangian state changes along the sequence of
stellar models.  }
\label{fig:M013rhotmulti}
\end{figure}

At model 12\,850, the luminosity generated in the hydrogen-burning
shell is still producing essentially the total stellar luminosity; at
this epoch, i.e. at about $t^\ast\approx$~-5034 yrs, the helium shell
contributes roughly one percent to the total luminosity. Within the
ensuing roughly 5000~yrs, the helium luminosity grows by roughly
eight orders of magnitude. When the temperature exceeds the limit for
the onset of $3\alpha$ burning off-center, the peak of the symmetric
temperature bump lies at about $0.174 \msol$ (see model number 12\,850
in Fig.~\ref{fig:M013rhotmulti}). The density at the temperature
maximum does not change significantly during the onset of the initial
flash (see the red line connecting $m=0.174 \msol$ in
Fig.~\ref{fig:M013rhotmulti}), all of the generated energy goes into
further rising the temperature.
\sidenote[][0.0cm]{This is exactly the
process associated with the thermal instability of thin
nuclear-burning shells in degenerate matter as encountered in
thermally pulsing AGB stars. The instability is well documented and
explained in standard textbooks on stellar
astrophysics \citep[e.g.][]{kw}}

Applying the energy equation to the steep temperature wall the
develops during the helium flash at the inner edge of this nuclear
burning shell
\begin{equation*}
  \partial_m L \approx - \cv \,\partial_t T \,,
\end{equation*}
the inward propagation speed of the wall can be estimated as
\begin{equation*}
\left\vert L \right \vert \tau \approx \cv \left\vert \frac{\Delta T}{\Delta
m} \right \vert
\left\vert \Delta m \right \vert^2 \,.
\end{equation*}
For a given extension in mass, $\Delta m$, and a given generated
luminosity, $L$, the magnitude of the temperature gradient, $\Delta T
/ \Delta m$, determines the timescale of propagation, $\tau$, of the
temperature pulse. 
\marginnote[-0.7cm]{The magnitude of $\tau$, to shift the temperature
  flank by its own mass thickness, varies between a few hundred
  seconds at flash peak (model 13\,025) and many thousand years once
  the helium-burning shell stabilizes again.}
Plugging numerical data from the evolutionary models into the above
estimate shows that the temperature wall does essentially not move in
mass during the initial flash.  Despite the low values of the opacity
\sidenote[][0.2cm]{Opacity ranges from about 0.1~cm$^2$/g close to the 
helium shell to about 0.01~cm$^2$/g in the strongly degenerate stellar center.}
in the inert stellar core, the prevailing temperature gradient is much
too low there to transport energy from the flash region to the
center. Therefore, only through the propagation of the temperature
wall towards the stellar center does energy penetrate the inert core.

\begin{marginfigure}
\includegraphics[width=1.08\textwidth]{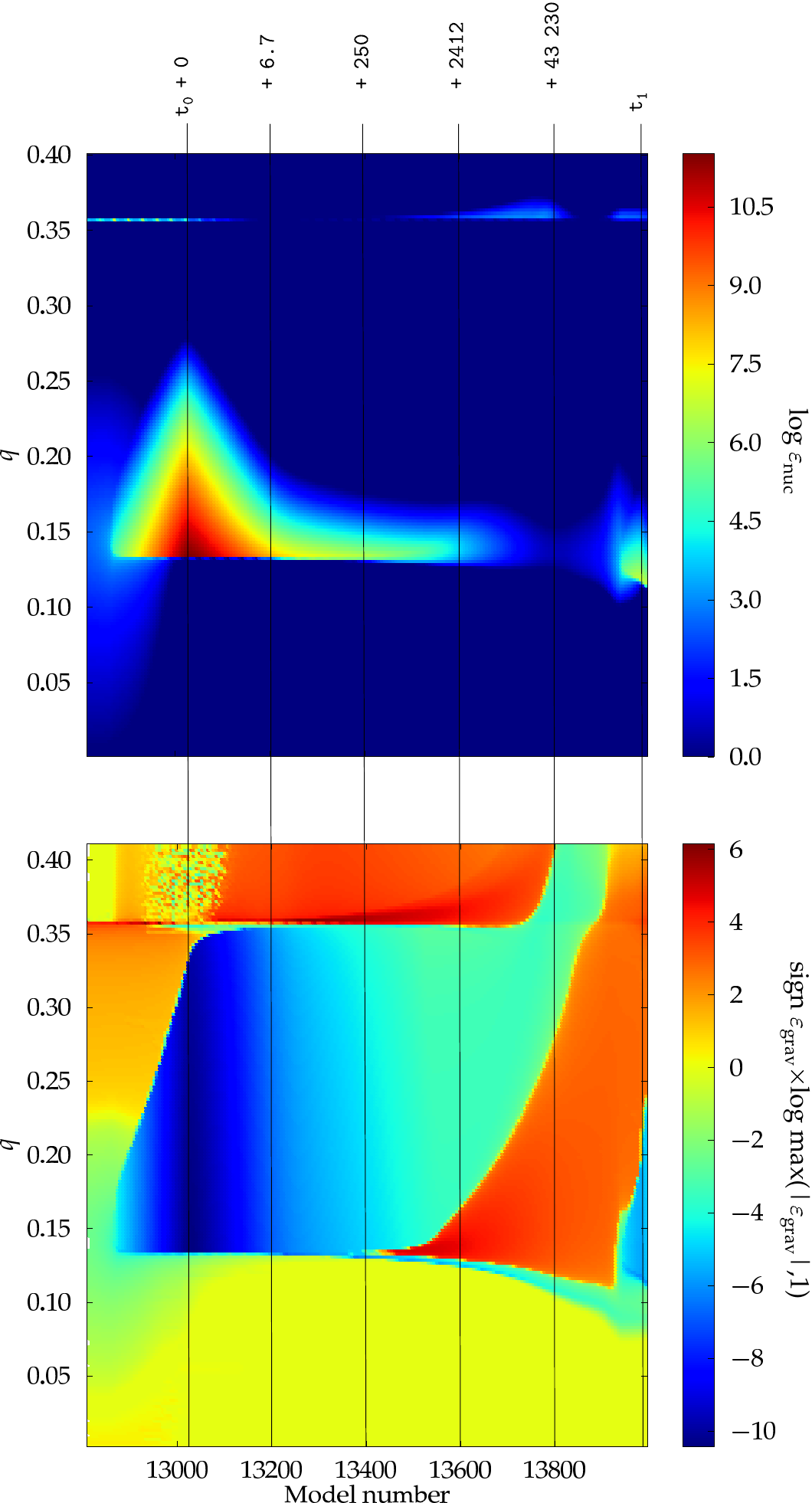}
\caption{Color-coded Kippenhahn diagrams showing the energetics in the
  deep stellar interior during the initial helium flash, which is
  parameterized on abscissa with the model number; this choice
  accentuates the fast evolution during the flash peak. Physical times
  in years, relative to the epoch of the initial flash, $t_0$, are
  overlaid over both panels. The epoch of the second helium flash is
  referred to as $t_1$. The ordinates are chosen to be the fractional
  stellar mass $q$. The top panel shows the nuclear energy generation
  rate $\varepsilon_\mathrm{nuc}/$erg/g/s and the lower panel a
  suitably transformed version of the gravitational energy generation
  $\varepsilon_\mathrm{grav}/$erg/g/s.  }
\label{fig:M013Kippi}
\end{marginfigure}

Once helium burning ignites off-center, the energy sink due to
neutrinos is quickly rendered irrelevant by the encountered magnitudes
of nuclear energy generation and the involved thermal energetics (see
also Fig.~\ref{fig:M013Kippi}). Hence, \emph{neutrino losses are essential
  to set the mass depth where helium burning ignites, but they do not
  influence the energetics of the helium flash itself.}

As the flash gains strength, convection sets in once a critical
temperature gradient, $\nabla T$, is exceeded on the outer flank of
the temperature bump. Since $\nabla T < 0$ on the lower-mass flank,
the stratification there is always stable against convection, hence
the peak becomes asymmetric. The inner edge of the He-shell convection
zone lies close to the maximum of the $3\alpha$ energy generation and
it extends well into the intershell region. At the phase of maximum
extension, at $t^\ast \approx 16$~yrs, the convection zone reaches out
to about $0.455 \msol$; this is still sufficiently below the inner
edge of the envelope convection zone (at $0.465 \msol$ as noted in
Fig.~\ref{fig:SchematicM013}), and separated by the hydrogen-burning
shell so that under `normal' burning conditions (i.e. PopI or PopII
conditions) no merging of the two respective convection zones and
hence no mixing was ever observed in simulations. The strong entropy
jump at the hydrogen-burning shell efficiently prevents the intershell
convection zone to advance too far out. Under special conditions,
which might not be realized in nature, a merging of convection zones
could be enforced \citep{Despain1976}. Also, modeling the evolution of
very low-metallicity (PopIII) stars revealed that ingestion of
hydrogen into the He-burning shell during the onset of core helium
burning can lead to enhanced convection zones in the core, due to the
additional onset of hydrogen burning, so that eventually the deepening
envelope convection zone can transport nuclearly processed material to
the surface \citep[e.g.][and references
therein]{Hollowell1990,Suda2010}.  During even later evolutionary
stages of any population-type star, the envelope convection zone
usually overlaps temporally with regions containing matter that was
previously modified by $3\alpha$ burning and mixes it
($2^{\mathrm{nd}}$ and $3^{\mathrm{rd}}$ dredge-up) into the
superficial layers of red giants along the AGB.

Figure~\ref{fig:M013Kippi} shows Kippenhahn diagrams for nuclear and
gravitational energy generation rates during the initial core helium
flash.  To provide a measure of the actual time evolution, epochs
measured in years elapsed since initial flash peak at $t_0$ are given
on top of the upper panel. The epoch of the second helium flash, at
model 13\,984, is denoted as $t_1=t_0 + 185\,218$~yrs.  The top panel
depicts the behavior of the hydrogen-burning shell at fractional mass
$q=0.35$ and the helium-burning shell with its basis at $q=0.13$. The
hydrogen shell is, compared with the helium shell, very narrow in mass
and goes essentially extinct shortly after the initial flash but
regains considerable strength after about 2000~yrs. Around minimum
radius of the model star (model 13\,833 at $t^\ast\approx
59\,400$~yrs) the hydrogen energy generation takes another dip because
the radius of the H-shell grows and cools as it partakes in the
expansion of the envelope towards the second thermal flash; shortly
before the second flash, after the H-shell contracted and heated
again, the hydrogen nuclear-energy generation regained its strength.

For a few years around the maximum of the initial flash, the
helium-burning shell achieves a mass depth of about $0.13 \msol$. From
$t^\ast=7$~yrs onward, the helium shell burns rather steadily but
with a tendency to get weaker and thinner.  The `gravitational energy
generation rate', $\varepsilon_\mathrm{grav}$, shown in the lower
panel is the temporal rate of change of the specific heat-content,
$Q$, of the stellar material: $\partial_t Q =
-\varepsilon_\mathrm{grav}$.  The locations of the helium- as well as
hydrogen-burning shell are clearly discernible at about $q=0.13$ and
$q=0.35$, respectively.  The stellar core: the sphere interior to the
He-shell does essentially not change its heat content,
i.e. $\partial_t Q = 0$, during the initial flash, which means that
state changes there are adiabatic (cf. red lines connecting e.g. $q=0$
and $q=0.16$ mass layers in Fig.~\ref{fig:M013rhotmulti}).

The envelope above the H-shell shows signs of contraction (red)
shortly after the initial He-flash when the intershell convection zone
reaches its maximum extension and essentially quenches the
H-shell. The intershell region is dominated by the convection zone
induced by the He-burning shell. From the scale on the color bar in
Fig.~\ref{fig:M013rhotmulti} we deduce that, centered around the
initial flash, $\varepsilon_\mathrm{grav}$ can easily compete in
magnitude with the nuclear counterpart from $3 \alpha$ burning; this
is the reason that stars with degenerate He-shell flashes are not
disrupted by the enormous nuclear energy input; it is roughly balanced
by decreasing the internal energy of the region and at the same time
expanding it in the deep gravitational well of the star. The shape of
the blue intershell region in the lower panel coincides with the
extension of intershell convection zone. Since the convective
time-scale is much shorter than the evolutionary time-scale after the
flash peak, the rapid propagation of physical information throughout
the intershell convection zone color along the $q$-axis (in particular
before about model 13\,600) levels out any developing
$\varepsilon_\mathrm{grav}$ gradient. The expansion of the envelope
(above the H-shell) is visible as the blue channel leaving the plot on
the upper right. On the other hand, the red region that establishes
itself on top of the He-shell around model 13\,400 and that continues
to extend outwards, also crossing essentially unimpeded the H-shell
before the second flash is dominated by an increase of the internal
energy but not by volume work, i.e. by contraction.

The intershell convection zone of the $1.3 \msol$ model sequence
reaches its maximum extension at about model no. 13\,200, i.e. at
$t^\ast\approx 6.7$~yrs after initial flash peak and it dies out at
$t^\ast \approx 3000$~yrs when $3\alpha$ burning develops a local
minimum after the central sphere below the helium-burning shell
expanded and cooled adiabatically (see Figs.~\ref{fig:M013Kippi} and
\ref{fig:M013Kippenhahn}). The presence of the intershell convection
zone causes the temperature gradient to be smoothed out, i.e. in the
convection zone it is considerably shallower than on the radiative
bottom-side of the He-burning shell (see model numbers $\ge$ 12\,950
in Fig.~\ref{fig:M013rhotmulti}); convection so deep in the stellar
interior is essentially adiabatic so that the stratification is isentropic 
there.
\sidenote{Since $\diff \ln T / \diff \ln P = \nabad$ obtains for the
  stratification, an ideal gas with
  negligible radiation pressure as well as a non-relativistic
  degenerate electron gas both support a slope 
\begin{equation*}
  \left.\frac{\diff \ln T}{\diff \ln\rho}
  \right\vert_{\mathrm{isentr.}} = \frac{2}{3} \,.
\end{equation*}
}
The steep temperature gradient on the bottom side of the
helium-burning shell persists during the whole flash and is initially
determined by the equation of state because the core cannot absorb any
significant amount of energy during the flash. Since only very little
mass is contained in the region of the sharp temperature drop
(referred to as \emph{temperature flank} in the following), the pressure
gradient remains comparatively flat so that
\begin{equation*}
\left\vert\frac{\diff \ln T}{\diff m}\right\vert \gg \left\vert\frac{\diff \ln
P}{\diff m}\right\vert  
\end{equation*}
obtains; the slope of the temperature flank on the $\rho - T$ plane
can then be approximated by
\begin{equation*}
\frac{\diff\ln T}{\diff\ln\rho} \approx \left( \frac{\partial \ln
T}{\partial \ln \rho}\right )_P = -\frac{1}{\delta}\,.
\end{equation*}
\marginnote[-0.8cm]{The notation of physical quantities in this
exposition is essentially that used in \citet{kw}; in particular the
equation of state is assumed to be of the functional form:
$\rho \sim P^\alpha\,T^{-\delta}\,\mu^\varphi$, with the
characteristic exponents $\alpha, \delta, \varphi$.} 
Hence, the higher the degeneracy around the temperature maximum, the
steeper the slope of the temperature flank. In the limit of
nonrelativistic full degeneracy, the inner temperature flank is
vertical in the $\log \rho - \log T$ plane.

During the early stage of the helium flash, the density of the core
region of the star does not change noticeably; i.e. the heating due to
the steeply increasing $3\alpha$ energy generation does not inflict
any significant expansion of the stellar material so that all
liberated energy goes into further rising the local temperature at the
nuclear burning shell. With increasing temperature in the flash
region, electron degeneracy diminishes and in conjunction expansion
can set in, i.e. an increasing fraction of the released nuclear energy
goes into volume work and therefore less energy remains for further
temperature rise. As an example: Since $\varepsilon_{3\alpha}\sim
\rho^2 T^{\,41\dots 19 \dots 12}$ at $T = 1\dots 2 \dots 3 \cdot
10^8$~K, the rapid decrease of the temperature dependence of the
energy generation rate with increasing temperature allows for an
increasingly smaller density reaction to compensate the energy
increase inflicted by some temperature rise: With the above numbers,
the energy release produced by a relative temperature increase $\Delta
T / T$ is neutralized by an associated relative density decrease of
$\Delta \rho / \rho = 20.5\dots 9.5\dots 6 \,\cdot\Delta T / T$. Once
such a limit is reached on the way of reducing the material's
degeneracy, the thermal flash has passed its maximum. In the $1.3
\msol$ model sequence, this happens when the He-burning shell reaches
the relatively high temperature of about $3\cdot10^8$~K, which lies
already slightly below $E_{\mathrm{F}}/kT=1$ for model number 13\,050
(see Fig.~\ref{fig:M013rhotmulti}).

After the energy generation in the He-shell saturates, the size of the
convection zone stalls too. The radiative regions overlying the
convection zone cannot carry away the surplus energy since the
evolution is still to rapid for diffusion to be effective. State
changes in the layers above the adiabatically stratified intershell
convection zone trace out loci with slopes very close to $2/3$ on the
$\log\rho - \log T$ plane
\sidenote{Introducing the constraint of an adiabatic state change
\begin{equation*}
\diff Q = 0 = \cp \diff T - \frac{\delta}{\rho}\diff P 
\end{equation*}
into the equation of state gives:
\begin{equation*}
\left.\frac{\diff \ln T}{\diff \ln \rho}\right\vert_{\diff Q=0} =
      \frac{\nabad}{\alpha - \nabad\cdot\delta} \,.
\end{equation*}
Both cases of relevance here, the ideal gas with negligible
radiation pressure with $\alpha=1,\,\delta=1,\,\nabad=2/5$ and the
non-relativistic electron degeneracy with $\alpha=3/5,\,\delta=0,\,
\nabad=2/5$, admit of 
\begin{equation*}
\left.\frac{\diff \ln T}{\diff \ln \rho}\right\vert_{\diff Q=0} =
      \frac{2}{3} \,.
\end{equation*}
}
(see e.g. the state changes at $q = 0.462$ in
Fig.~\ref{fig:M013rhotmulti}). Only after about model number 13\,400
enough time has elapsed for diffusion of radiation to influence state
changes in the intershell region (see again the state change at mass
shell $q=0.462$ just before model 13\,500 in
Fig.~\ref{fig:M013rhotmulti}) so that their loci on the $\log\rho -
\log T$ deviate from lines with slope 2/3.

The computations show that the expansion speed of the intershell
region above the helium flash proceeds as $\partial_m v \approx
\mathrm{const.}$; using this in the continuity equation together with
the assumption of adiabatic state changes, we find
\begin{equation*}
\frac{\partial_t T}{T} \sim \frac{1}{r}\,.
\end{equation*} 
Hence, in the inner parts of the radiative regions of the inter-shell
layers cool stronger than the higher lying ones: This can be observed
in Fig.~\ref{fig:M013rhotmulti} to the left of $E_{\mathrm{F}}/kT =
1$, just outside of the intershell convection zone where a positive
temperature gradient develops during the initial flash cycle.

In the degenerate core, state changes are also adiabatic (see lines of
$q=0$ and $q=0.16$ in Fig.~\ref{fig:M013rhotmulti} or lower panel of
Fig.~\ref{fig:M013Kippi}). The expansion of the He-flash domain lifts
its matter into regions of lower gravitational acceleration; hence,
this expansion reduces the pressure on the inner core so that the
very core expands adiabatically. Expansion is largest where degeneracy
is lowest, i.e. around the He-flash shell. Due to near-adiabaticity of
the state change, temperature change is biggest where expansion is
biggest. Therefore, the originally slightly positive temperature
gradient goes negative at the inner edge of the helium burning zone.


\newthought{The H-shell essentially switches off} temporarily at about
$t^\ast = 7$~yrs when the adiabatic expansion of the intershell
region associated with the He-flash has sufficiently reduced the
temperature at the H-shell. But only at $t^\ast \approx$ 44~yrs has
the surface luminosity dropped by $\vert\Delta L_\ast/L_\ast\vert =
0.1$~--~this being taken as the sign that the star starts to leave the
top of the giant branch at 2417~$L_\odot$ (cf. column 4 in
Table~\ref{tab:TP1Selecta}). At the end of the initial flash, the
star's luminosity reached 113~$L_\odot$, which is roughly twice the
luminosity at which it will finally settle to centrally burn
helium. The core, being here the fractional stellar sphere lying
inside of the H-shell, is hydrostatically insulated from the envelope,
so that the core can change mechanically \emph{without} dragging along
the envelope \citep[e.g.][]{Stein1966, Sugimoto1980}.  Therefore, the
delay of the surface luminosity reaction to what happens in the deep
interior can be attributed to the Kelvin-Helmholtz time of the
\emph{radiative} intershell layers which must be overcome by photon
diffusion. Once the envelope adjusts to the new energetic situation,
the bottom of the envelope convection zone retreats. By $t^\ast =$
2412~yrs, i.e. model 13\,600, helium burning eases off sufficiently
for the He-shell convection zone to disappear
(cf. Fig.~\ref{fig:M013Kippi}).  Around model number 13\,740 ($t^\ast
\approx 17\,650$~yrs), the first material layers of the contracting
envelope hit and reflect on the steep density gradient at the
H-burning shell.  The generated `pulse'
\sidenote{ The trace of the outgoing pulse is seen as the narrow blue
  locus close to the upper right corner of the plot on lower panel of
  Fig.~\ref{fig:M013Kippi}.
}
propagates back into the envelope and defines, upon its arrival at the
stellar surface, the minimum radius and accordingly the minimum
luminosity reached during the initial pulse ($47 \lsol$ at model
no. 13\,834, i.e. at $t^\ast \approx 62\,000$~yrs). The epoch of
pulse generation at the H-shell goes along with a re-expansion and
cooling of the H-shell and induces a phase of a nearly extinct
hydrogen-burning shell. The spatial movement of the H-shell is
reminiscent of a suction reaction inflicted by the rapid expansion of
the overlying envelope.  Meanwhile, the layers below the H-shell
contract adiabatically (to rise density and temperature along loci of
$\diff \ln T / \diff \ln \rho =2/3$~--~cf. 
Fig.~\ref{fig:M013rhotmulti}) to eventually lead to the
next thermal instability of the helium-burning shell.
  
\section{The thermal pulsing episode of the helium shell}
\label{sect:tps}
The initial helium flash is only the first of a series of flashes that
accompany the inward burning of the He-shell toward the stellar
center. Depending, among other parameters, on the stellar mass, the
number of ensuing flashes can exceed a dozen at low stellar masses;
the number goes down as the stellar mass increases and the initial
flash takes place closer (measured in mass) to the stellar center. The
physics of the thermal stability of thin nuclear-burning
shells is well understood and documented e.g. in \citet{kw}. Most
frequently, shell instability is encountered and studied in the
evolution of stars along the asymptotic giant branch. The physics of
the instability prevailing in the inward-burning He-shell at the onset
of core helium burning of low-mass stars is, however, the same.

\begin{figure}
\centering
\includegraphics[width=1.0\textwidth]{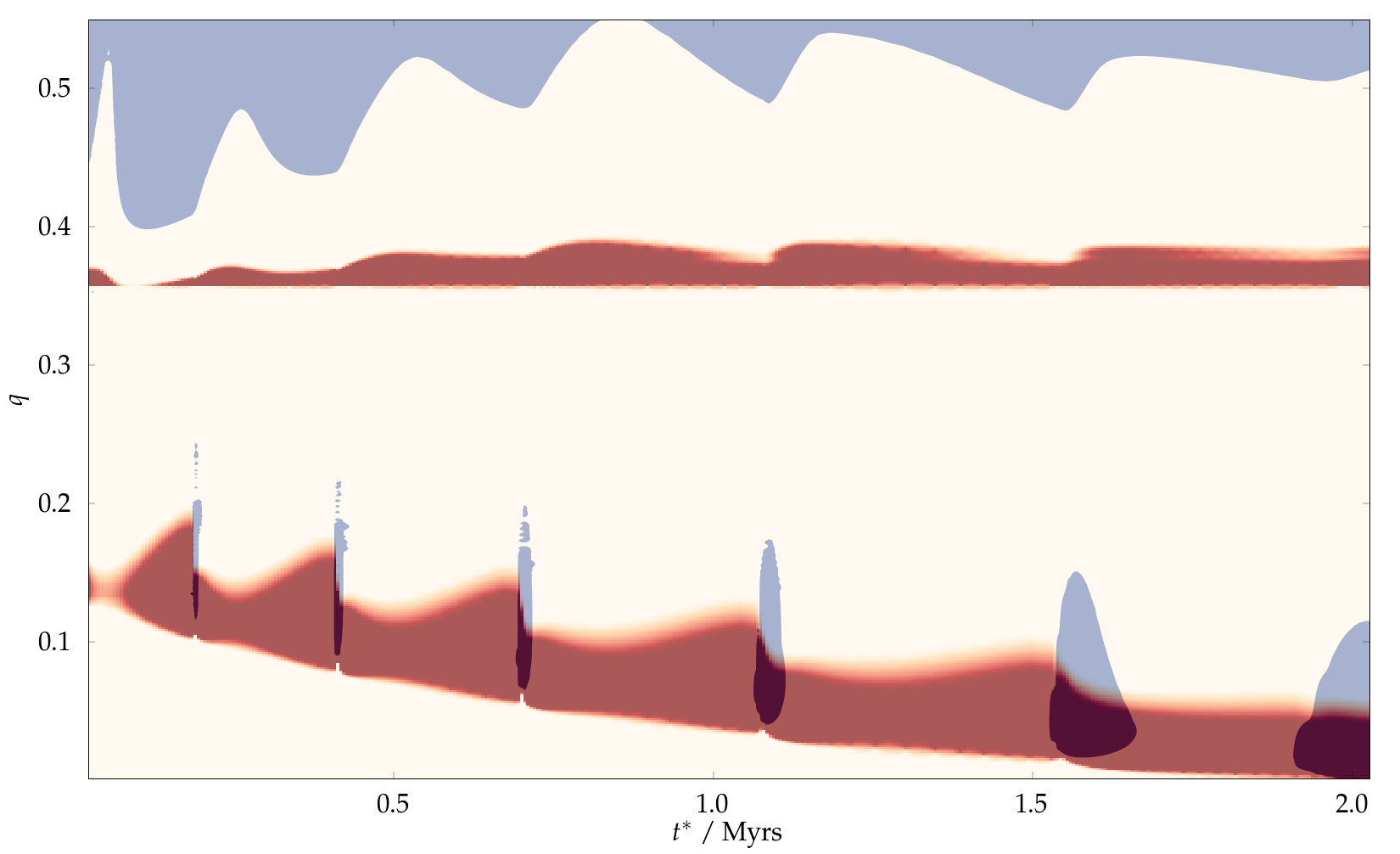}
\caption{Kippenhahn diagram for the inner $54 \%$ of the $1.3 \msol$
model star's mass, covering the end of the first (starting at model
no. 13\,400) and the ensuing five thermal flashes until central helium burning
takes the star into a quasi-static state again. The time
is measured in mega-years (Myrs) relative to $t_0$. Regions with
$\varepsilon_\mathrm{nuc} > 10$~erg/g/s are colored in red. Convective
regions, as prevailing according to Schwarzschild's criterion, are
shaded in light blue. Notice that, according to our models, the bottom
boundary of the envelope convection zone never penetrated deep enough
into the star after the initial helium flash so that \emph{no}
dredge-up of $3\alpha$-processed material takes place. 

\smallskip

\noindent
The spotty character of the He-shell convection zone is spurious,
caused by the Delaunay triangulation of strongly unequally spaced
stellar-evolution data onto the regular grid that was computed for the
color-coded plot.}
\label{fig:M013Kippenhahn}
\end{figure}

In all cases known to us, the first flash of the onset of helium
burning in a low-mass star's core is by a large margin the most
energetic one. In the $1.3\,\msol$ sequence exemplified here, the
helium luminosity grows to $10^{9.25} \lsol$ during the initial flash.
During the second flash, the helium luminosity is already about five
orders of magnitude weaker with the tendency of a continuing weakening
of the later flashes (see Fig.~\ref{fig:M013lums}).
The temporal evolution during the thermal pulsing phase of the nuclear
burning ($\varepsilon_\mathrm{nuc} > 10$~erg/g/s plotted in red) in
the stellar core is displayed in Fig.~\ref{fig:M013Kippenhahn}; the
He-shell convection zone as well as the bottom of the envelope
convection zone~--~both plotted in blue~--~are added for better
orientation.  The initial pulse (shown in an appropriately scaled way
in Fig.~\ref{fig:M013Kippi}) lies just off to the left of the diagram.
Pulses two to six can be easily identified by the He-shell convection
zone which they trigger.

The last pulse, which directly leads to sustained central helium
burning is visible on the right of Fig.~\ref{fig:M013Kippenhahn}. The
initial rise of helium burning looks like the previous \emph{shell}
instabilities; however, later during the flash, as the shell burns into
the star's center it can not react as before, the shell morphs into a
central instability and hence into its rising the temperature at
essentially constant density.  

The innermost part of the envelope convection zone is visible along
the top of the figure as the wavy blue band. The bottom of the
convection zone reacts on the nuclear activity in the stellar
core. Just as during the initial flash, also during the later ones the
inner edge of the convective envelope recedes after each helium flash
to re-penetrate deeper into the star during the later part of the
thermal-pulse cycle. The bottom of the convective envelope never
reached into material that was nuclearly processed during the thermal
pulses so that there is, in our computations, no chance of dredging-up
$3\alpha$ burning products.  While the helium shell burns its way into
the stellar center on the timescale of about 2 Myrs, the hydrogen
shell remains essentially at constant mass depth. As can be deduced
from Fig.~\ref{fig:M013Kippenhahn}, the H-burning shell reacts to the
thermal pulsing He-shell by cyclically growing fatter shortly after
each He-flash to later-on thin out again.

\begin{marginfigure}[-1.5cm]
\includegraphics{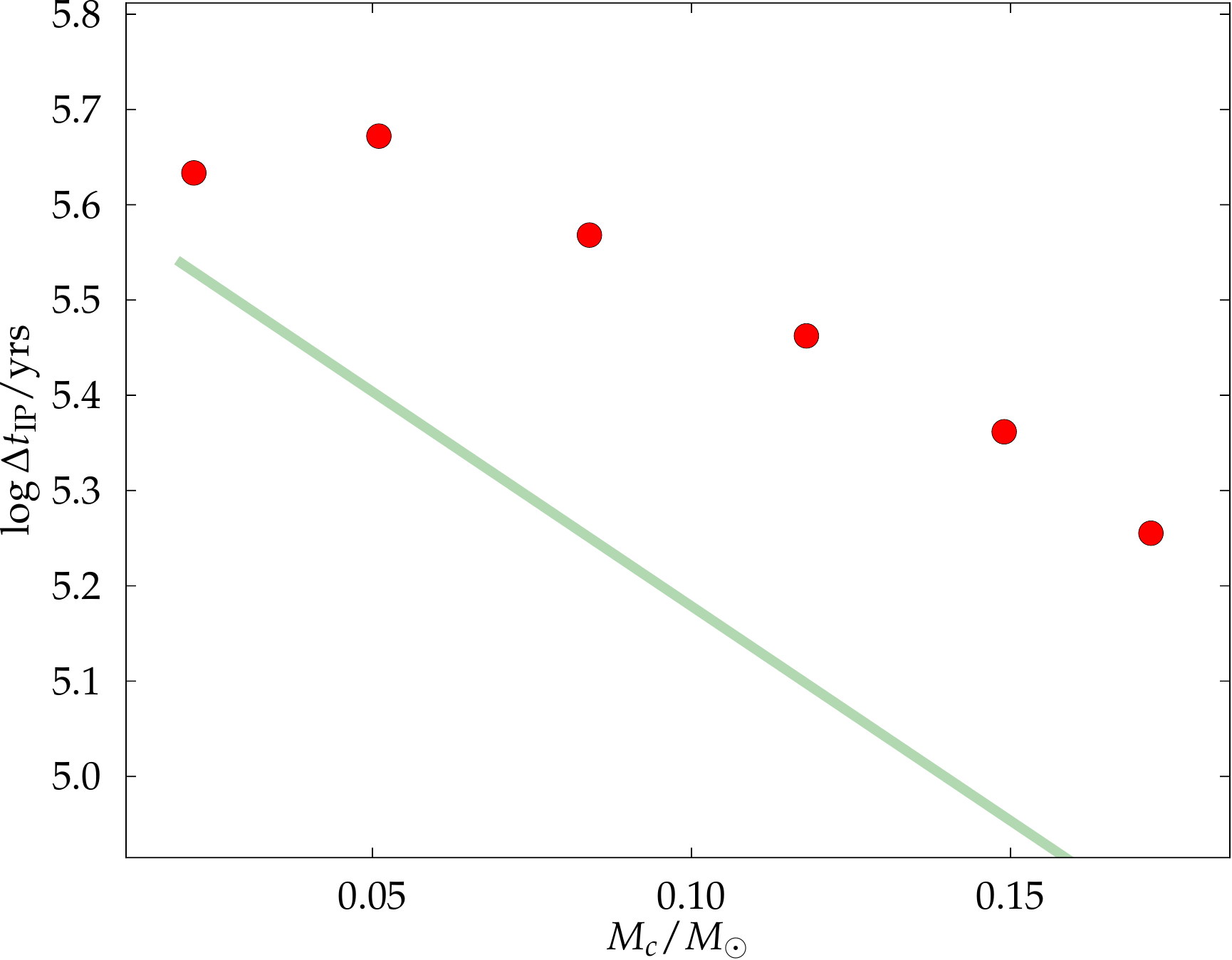}
\caption{Interpulse-period~--~core-mass relation as observed in the
evolution computations of the $1.3 \msol$ models (red points),
compared with the analytical fit by \citet{Despain1981}: 
$
\log \Delta t_{\mathrm{IP}}/\mathrm{yrs} = 
            5.629 - 4.504\left({m_{\mathrm{c}}}/{\msol}\right)\,
$
which is shown as full line in the figure.}
\label{fig:DeltatIP}
\end{marginfigure}

Figure~\ref{fig:M013Kippenhahn} illustrates furthermore the regularity
of the thermal flash cycles. As for the stars along the AGB, the time
between two flashes, the interpulse period (IP), $\Delta
t_{\mathrm{IP}}$, appears to correlate with the core mass,
$m_{\mathrm{c}}$.  For the $1.3 \msol$ model sequence highlighted
here, the core-mass~--~interpulse period relation is shown 
by dots in Fig.~\ref{fig:DeltatIP}; they can be fitted by the relation
\begin{equation*}
\log \Delta t_{\mathrm{IP}}/\mathrm{yrs} = 
            5.83 - 3.23\left(\frac{m_{\mathrm{c}}}{\msol}\right)\,.
\end{equation*}
The core mass, $m_{\mathrm{c}}$, was measured at the maximum of
$3\alpha$ burning, the interpulse period, $\Delta t_{\mathrm{IP}}$, be
the time passed between two He-flash peaks. The last point in the
core-mass~--~interpulse-period diagram, i.e. the one belonging to the
last helium flash, which started essentially centrally, was excluded
from the analytical fit. The last flash is no longer pure in the sense
that it is distorted by its proximity to the stellar center. Indeed,
central burning evolves continually out of this distorted last
flash. For comparison, the analytical fit to the interpulse-periods of
the $0.6 \msol$ models of \citet{Despain1981} is shown as the
continuous line in Fig.~\ref{fig:DeltatIP}.

As the last helium flash `hits' the star's center, the
$3\alpha$-burning matter cannot cool anymore by adiabatic expansion
induced by the underlying material as hitherto, but it stays initially
at roughly constant density while the temperature rises so that the
material's degeneracy is being reduced. Once degeneracy is low enough,
the density starts to diminish. All this happens, as illustrated in
\citet{Paxton2011}, at essentially constant total pressure; this is
achieved by the lack of any expansion/contraction of the material
overlying the helium-burning core.  Once, the stellar center's
degeneracy is lifted, central helium burning proceeds in complete
equilibrium under essentially ideal-gas conditions.

\begin{figure}
\includegraphics{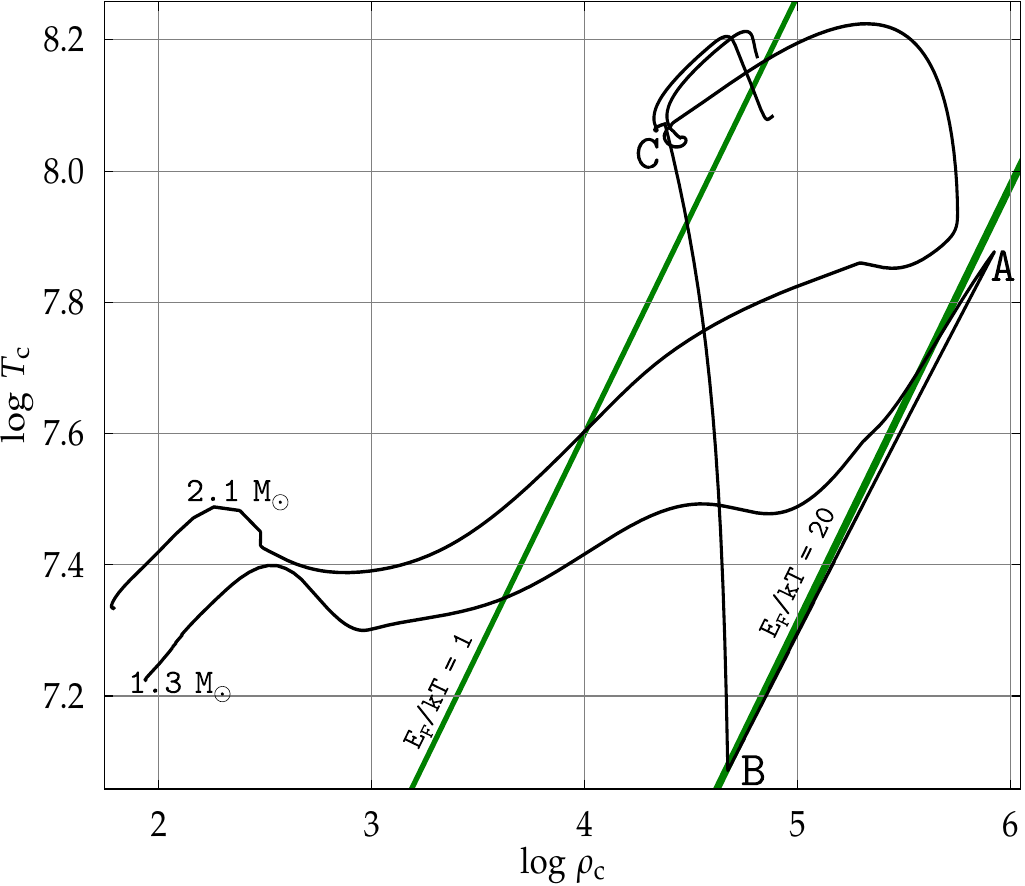}
\caption{Evolution on the logarithmic density~--~temperature plane of
  the centers of the model stars shown in Fig.\ref{fig:HRD}.
  Epochs A, B, and C along the locus traced out by the center of
  the 1.3~$\msol$ model star define the edges of the `wedge', which is
  referred to in the text.  In the case of the 2.1~$\msol$ star whose
  helium burning starts in the center, the corresponding density stays
  initially essentially constant but the temperature rises strongly and
  maneuvers the star's center out of degeneracy already during
  the initial flash.}
\label{fig:M013M021rhoctc}
\end{figure}
\newthought{The major features of the wedge} as traced out on the
$\log\rhoc - \log \tc$ plane by a low-mass star's center during the
onset of core helium burning can eventually be understood
qualitatively: In Fig.~\ref{fig:M013M021rhoctc}, epoch A indicates the
tip of the first giant branch when the initial flash sets in. Epoch B
marks the arrival of helium burning in the star's center. Finally,
state C is reached when the final thermal flash removes electron
degeneracy in the center and quiescent \emph{central} helium burning
takes over. Notice that evolution of the star's center from A to B on
the $\log\rhoc - \log \tc$ does not proceed monotonously. During each
thermal flash the star's center moves back and forth along a locus of
essentially $E_{\mathrm{F}}/kT = \mathrm{const.}$; the magnitude of
$E_{\mathrm{F}}/kT$ which prevails in a star's center remains
essentially constant, determined by the total stellar mass.  During
the luminosity decline of each pulse cycle, the central density (and
temperature) decline; during the ascending phases the cycles,
luminosity and central density (and also temperature) rise again
slightly. The magnitudes of the density rises are smaller than the
cyclic declines, so that an effective reduction in central density and
temperature result throughout the flash cycles.  The fact that
degeneracy of the stellar center does hardly change during the series
of thermal pulses is attributable to lines of constant degeneracy and
loci traced out by adiabatic state changes, both having the same
slope.

\begin{figure}
\includegraphics{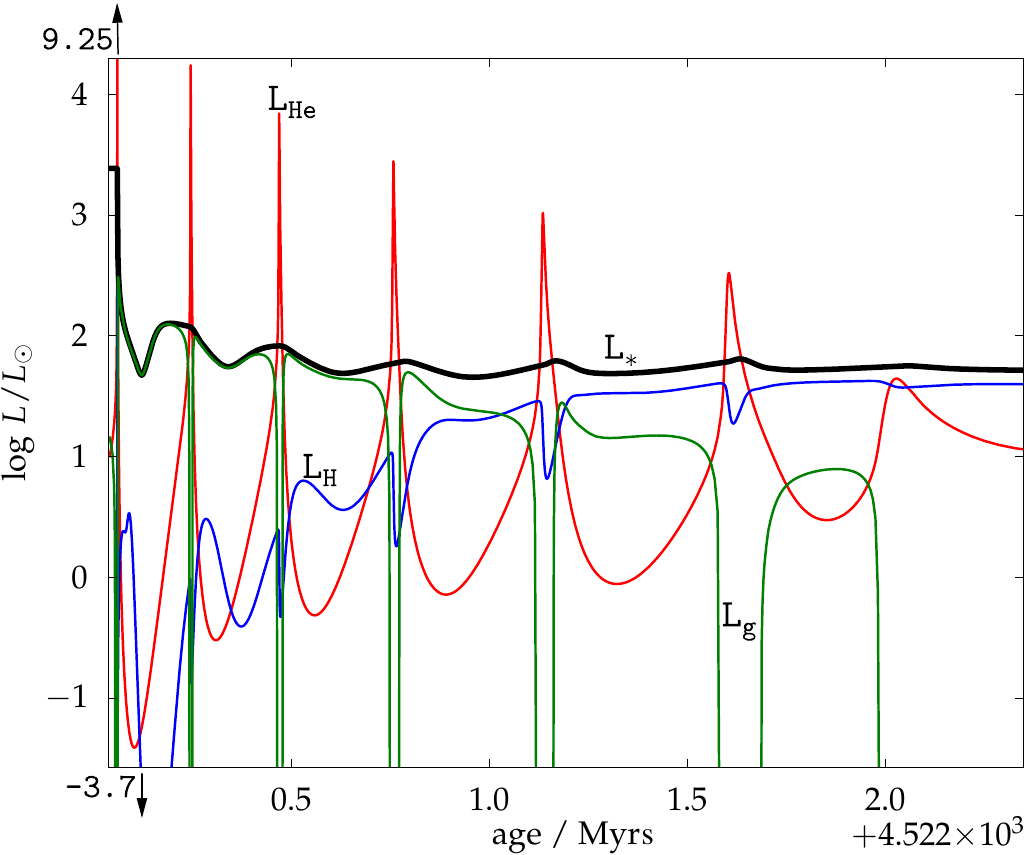}
\caption{Evolution of the total stellar luminosity ($L_{\ast}$, heavy
  line) and its constituent components, luminosity from hydrogen
  burning ($L_{\mathrm H}$), luminosity from helium burning
  ($L_\mathrm {He}$), and the `gravitational luminosity' ($L_{\mathrm
    g}$) during the onset of helium burning in the $1.3 \msol$ star
  model.}
\label{fig:M013lums}
\end{figure}

\newthought{Even though the helium luminosity dominates} early phases
of the initial flash cycle, other components, usually not being in the
limelight of attention, affect the star's luminosity once the thermal
instability of the helium shell fades away. 
Figure~\ref{fig:M013lums} shows that it is the gravitational luminosity
\sidenote{ The casually referred to \emph{gravitational luminosity} is
  defined as 
\begin{equation*}
 L_{\mathrm{g}} \doteq \int \varepsilon_{\mathrm{grav}}\, \diff m \,; 
\end{equation*}
it is a measure of a star's departure from thermal equilibrium.
}, 
$L_{\mathrm{g}}$, which dominates all other luminosity contributions
(except for $L_\mathrm {He}$ during the short flash peaks) for the
first roughly $1$~Myr after the initial flash; during this phase, the
$1.3 \msol$ star evolves from the tip of the giant branch to its clump
position at about $60 \lsol$. The gravitational luminosity takes over
the r\^ole of the dominant luminosity source at around
$t^\ast=2400$~yrs, after the $L_\mathrm {H}$ minimum passed and the
envelope shrunk from 160 to $43 \rsol$. During this later phase of the
initial pulse cycle it is the envelope above the H-burning shell
that contributes most to $L_{\mathrm{g}}$. The biggest positive
$L_{\mathrm{g}}$ contribution comes from the very base of the
envelope, the region closest to the H-burning shell, which lies also
in the steepest part of the gravitational potential.
\sidenote[][-2.0cm]{The thermal time-scale ($\int\cv T \diff
  m/L_\ast$) of the envelope, extending from the photosphere to the
  outer edge of the H-burning shell, computed for the conditions at
  the tip of the red-giant branch amounts to a few hundred years. Only
  when assuming the envelope to encompass also the intershell region
  could the corresponding thermal heat content power the star for
  about $10^4$~yrs.}
The $L_{\mathrm{g}}$ behavior during the subsequent flashes, with
successively lower amplitude and with accordingly smaller radius and
thermal variation, is qualitatively the same as that during the
initial He-flash cycle. 

After the initial flash, the hydrogen shell dims out so that at model
13\,200 (i.e. at $t^\ast=6.7$~yrs) it reaches a minimum at $L_\mathrm
{H}=10^{-3.7}\lsol$ as also seen in
Fig.~\ref{fig:M013Kippi}. Figure~\ref{fig:M013lums}, on the other
hand, shows the hydrogen shell to regain sufficient strength after the
forth pulse and overtaking $L_{\mathrm{g}}$; the hydrogen-luminosity
continues to grow and finally takes over as the dominant nuclear
energy source of the star during the ensuing inter-flash phases. Even
after central helium burning is established, the H-shell retains the
status as the dominant nuclear energy source. At most, the $3\alpha$
luminosity exceeds the H-luminosity by 35 \% during the initial peak
of central helium burning, but most of the time it is clearly
superseded by the energy output of the H-burning shell.
\sidenote[][-0.2cm]{$L_{\mathrm{H}}/L_{\mathrm{He}}=2.4 - 3.4$ during most of
  the central He-burning.}

During the whole thermal-pulse phase, however,
$\varepsilon_{\mathrm{grav}} \approx 0$ obtains in the central sphere
bounded by the He-shell, i.e. it stays inert in every energetic
respect. It is the diffusion time-scale of the He-burning shell into
this inert degenerate central sphere which determines the duration of
the thermal pulsing phase. Analytical guesswork is cumbersome due to
the rather strong variation of the pertinent physical quantities at
the He-burning shell during the flash cycles; depending on the
particular choices, durations between $10^5$ and $10^7$~yrs
result; stellar evolution computations yield $2\times 10^6$~yrs.

\begin{marginfigure}[-0.0cm]
\includegraphics[width=0.95\textwidth]{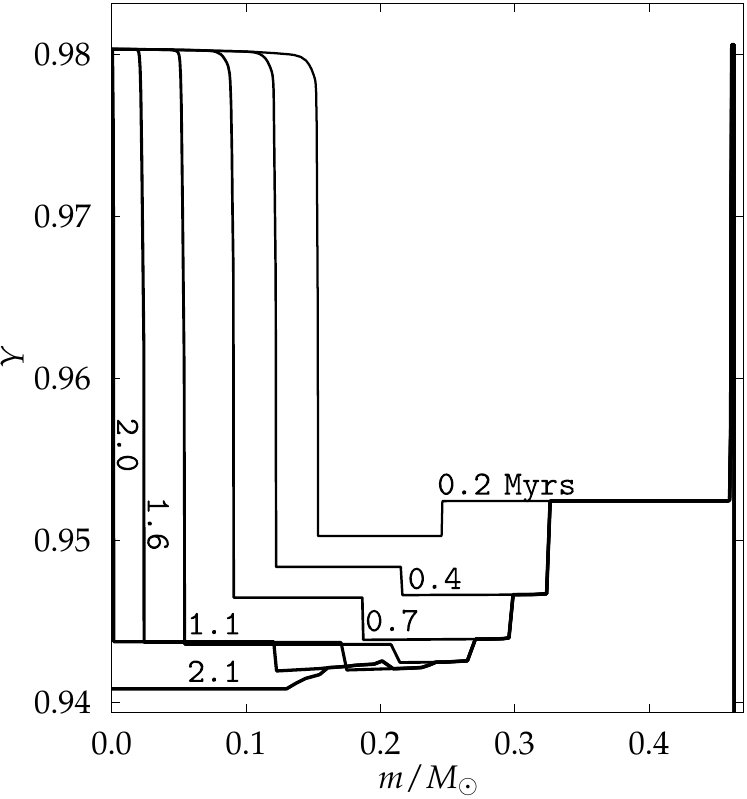}
\caption{Spatial variation of the helium abundance in the cores of
  $1.3 \msol$ models. The epochs of the snapshots, measured in Myrs as
  relative times $t^\ast$, are labeled on the respective
  profiles. The lowest six epochs were chosen at the local minima of
  $L_\mathrm{He}$ between the thermal pulses. At $t^\ast=2.1$~Myrs,
  central helium burning was already in progress.}
\label{fig:HeCore}
\end{marginfigure}

\newthought{Off-center helium flashes} burn some of the helium (see
Fig.~\ref{fig:HeCore}) mainly to carbon and oxygen so that heavier
material overlies lighter one in the core~--~a potentially unstable
situation. Since stellar material is not isothermal in the relevant
regions, the stratification is not just Rayleigh-Taylor unstable; the
prevailing temperature profile can stabilize a certain magnitude of
molecular-weight contrast. The non-vanishing diffusivity of heat
requires then the stability condition against double-diffusive mixing
to be
studied.\sidenote[][1.0cm]{This is the same physical phenomenon as the
  salinity steps observed in the stratification of sea water
  observable at favorable places on the globe. Therefore, even in
  highly compressible stellar astrophysics, this double-diffusive
  instability is mostly referred to as salt-finger or thermohaline
  instability~--~the names used in oceanography.}
Even though the qualitative picture of the onset of core-helium
burning as observed through the results from stellar structure and
evolution 
computations\sidenote[][+1.0cm]{Essentially measured heuristically by comparing
  the results of different generations of stellar-evolution codes, all
  with different micro-physics and many of them with different
  numerics.}
seems to be robust, any quantitative study requesting information on
abundances and abundance profiles, either as observed directly on
red-giants' surfaces or possibly deduced via observed oscillation
frequencies of red giants will require detailed multi-dimensional CFD
simulations of the stellar core region. The same applies to the study
of the detailed influence of the dynamical convection and what happens
at the corresponding convective boundaries at the He-burning shell
during the initial flash. 

\newthought{Acknowledgment} NASA's Astrophysics Data System was used
extensively for this exposition. Without the open-source project MESA,
none of the illustrations and analyses presented in this paper would
have been possible; the efforts and willingness to communicate, in
particular of the `chief developer', Bill Paxton, are highly
appreciated and equally admired. I am grateful to Hideyuki~Saio who
helped to improve the content of this exposition by critically
commenting on the typescript. H.~Harzenmoser stimulated and followed
closely also this project; he tried to keep up the author's spirits
during numerous culinary late-night sit-ins that reverberated from
exegeses on the virtues of fostering understanding rather than
aggregating yet more information.

\bibliographystyle{aa} 
\bibliography{StarBase}

\end{document}